\documentclass[conference,10pt]{IEEEtran}
\usepackage{epsfig,epsf,rotating,setspace,latexsym,amsmath,amssymb,amsfonts,bm,theorem,subfigure,epstopdf}
\usepackage{cite,authblk}
\usepackage{bbm}
\usepackage{algorithm}
\usepackage[noend]{algpseudocode}
\usepackage{color}
\usepackage{mathtools}
\usepackage{soul}

\algrenewcommand\algorithmicforall{\textbf{foreach}}
\algrenewcommand\algorithmicindent{.8em}

\IEEEoverridecommandlockouts
\allowdisplaybreaks

\begin{document}
 
%\title{Information Reliability in Age-Sensitive Gossip Networks }
% \title{Reliable and Unreliable Sources in Age-Sensitive Gossip Networks}
\title{Reliable and Unreliable Sources in \\ Age-Based Gossiping}
%\title{Reliable and Unreliable Sources in Age-Sensitive Gossiping}
 
\author{Priyanka Kaswan \qquad Sennur Ulukus\\
        \normalsize Department of Electrical and Computer Engineering\\
        \normalsize University of Maryland, College Park, MD 20742\\
        \normalsize  \emph{pkaswan@umd.edu} \qquad \emph{ulukus@umd.edu}}
 
\maketitle

\begin{abstract}
We consider a network consisting of $n$ nodes that aim to track a continually updating process or event. To disseminate updates about the event to the network, two sources are available, such that information obtained from one source is considered more reliable than the other source. The nodes wish to have access to information about the event that is not only latest but also more reliable, and prefer a reliable packet over an unreliable packet even when the former is a bit outdated with respect to the latter. We study how such preference affects the fraction of users with reliable information in the network and their version age of information. We derive the analytical equations to characterize the two quantities, long-term expected fraction of nodes with reliable packets and their long-term expected version age using stochastic hybrid systems (SHS) modelling and study their properties. We also compare these results with the case where nodes give more preference to freshness of information than its reliability. Finally we show simulation results to verify the theoretical results and shed further light on behavior of above quantities with respect to dependent variables.

\end{abstract}

\section{Introduction}\label{sec:intro}

We consider a system where a set of $n$ nodes wish to track an event or a process (E), which gets updated according to a Poisson process with rate $\lambda_E$. However, to transmit the information about the event to the nodes, two sources are available, such that one of these sources is more reliable than the other and is expected to transmit more accurate information. We call the former as \emph{reliable} (R) source and the latter as \emph{unreliable} (U) source, and they send out packets to the network with total update rate of $\lambda_R$ and $\lambda_U$, respectively. This setting could arise when multiple sensors are monitoring physical environment and send updates to an IoT network, such that some of these sensors are unreliable. Another example could be when multiple websites or news sources are delivering scores of a sports event to a group of interested viewers, such that some of these sources are more reliable than others. 

We notice that in the above settings, the event of interest is dynamic in nature. In case of time-sensitive dynamic information, network nodes are usually interested in obtaining the latest possible information, which can be quantified by timeliness metrics such as age of information \cite{Kosta17agesurvey, Sun19agesurvey, yates21agesurvey,Yates21gossip_traditional}, version age of information \cite{Yates21gossip,kaswan22jammerring}, binary freshness \cite{bastopcu2020LineNetwork,kaswan_isit2021,Bastopcu21gossip}, age of incorrect information \cite{maatouk_age_incorrect}, etc, that are commonly used in literature. In this work we employ version age of information metric as we associate a version number to every information about the event. If $V_E(t)$ denotes the version number corresponding to the current state of event and $V_i(t)$ denotes the version number of the information about the event present at node $i$, then the instantaneous version age of information at node $i$ is defined as $X_i(t)=V_E(t)-V_i(t)$, where $V_E(t)$ increments by one every time the event gets updated.

\begin{figure}[t]
\centerline{\includegraphics[width=0.8\linewidth]{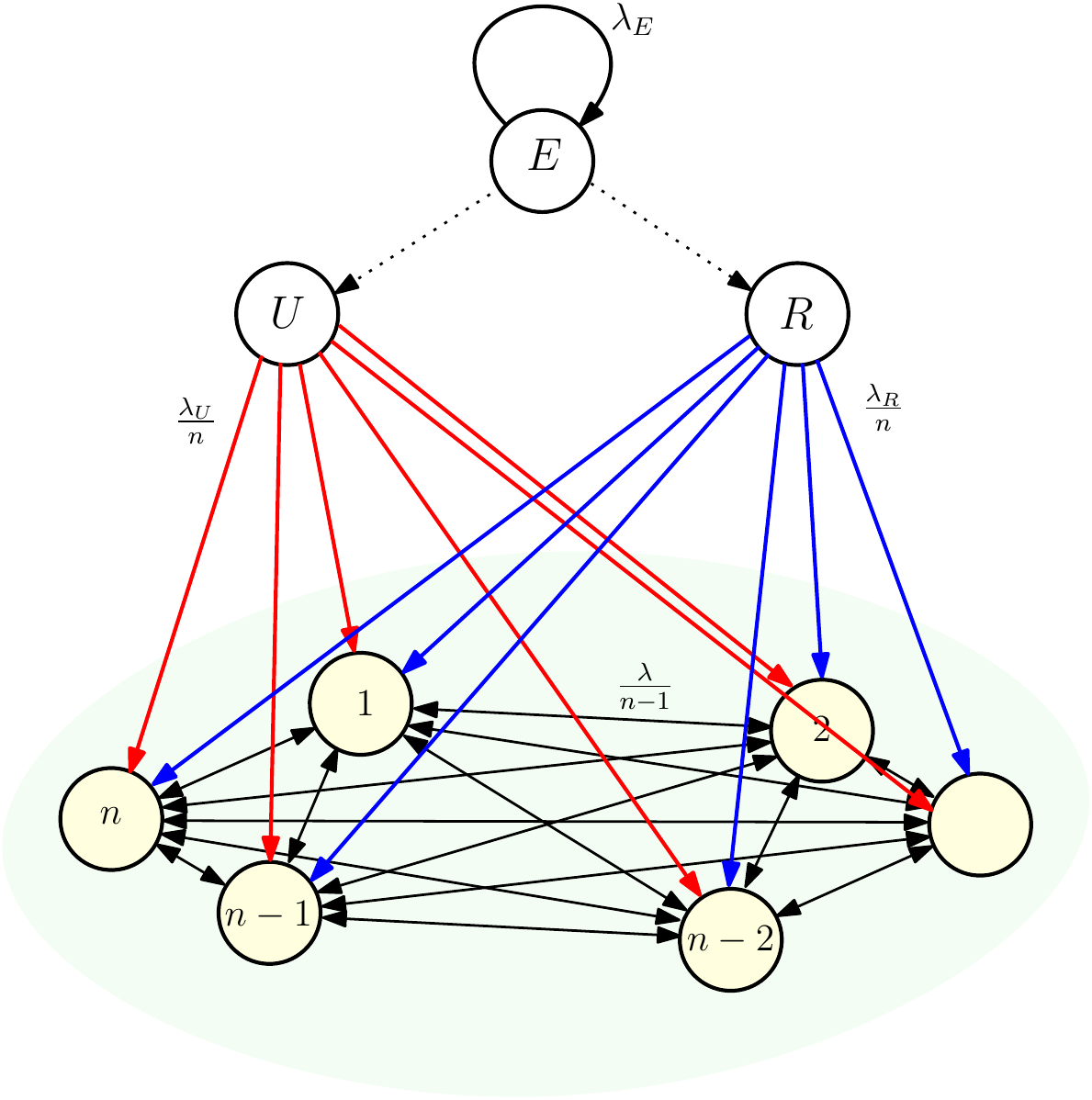}}
\caption{System model with reliable (R) and unreliable (U) sources observing an event (E) that updates with rate $\lambda_E$. Reliable (R) and unreliable (U) sources disseminate the information to the network at total rates $\lambda_R$ and $\lambda_U$. Each user gossips with its neighbors with rate $\lambda$ over a fully-connected network.}
\label{fig:system_model}
\vspace*{-0.4cm}
\end{figure}

In addition to the sources disseminating updates to the network, the nodes in the network further aim to expedite dissemination to improve their version age of information and information reliability through gossiping \cite{Yates21gossip_traditional,Demers1987EpidemicAF-short, Minsky02cornellthesis, vocking2000, Pittel1987OnSA, deb2006AlgebraicGossip, devavrat2006, Sanghavi2007GossipFileSplit, amazondynamo-short, Cassandra, Yates21gossip, baturalp21comm_struc, Bastopcu21gossip, kaswan22slicingcoding,mitra_allerton2022,bastopcu_agent_gossip}. Gossip algorithms are decentralized algorithms where nodes randomly contact their neighbors to exchange packets, and networks employing gossip protocols have been widely studied in the literature from dissemination time perspective \cite{vocking2000,deb2006AlgebraicGossip,devavrat2006,Sanghavi2007GossipFileSplit} and timeliness perspective \cite{Yates21gossip,Yates21gossip_traditional, baturalp21comm_struc, Bastopcu21gossip, kaswan22slicingcoding,kaswan22jammerring,kaswan22timestomping}.

The works most closely related to this paper are \cite{Yates21gossip,Yates21gossip_traditional, baturalp21comm_struc, Bastopcu21gossip, kaswan22slicingcoding,kaswan22jammerring,kaswan22timestomping}. \cite{Yates21gossip_traditional, Yates21gossip} derive the recursive linear equations using stochastic hybrid system (SHS) framework for expected age and expected version age, respectively, \cite{baturalp21comm_struc} studies the expected version age in clustered gossip networks,  \cite{Bastopcu21gossip} provides analogous results for binary freshness metric, \cite{kaswan22slicingcoding} attempts to improve age bounds in gossip networks using file slicing and network coding, and \cite{kaswan22jammerring} and \cite{kaswan22timestomping} study the effects of jamming and timestomping adversaries on gossip networks. All these works have a single source of information responsible for transmitting latest information packets to the network, and the nodes exchange packets with the single goal of improving their freshness of information.

In this paper, we consider two kinds of sources, a reliable source and an unreliable source, who are always assumed to have the latest information about the event, resulting in their respective version age of information $X_R(t)$ and $X_U(t)$ to be zero at all times. The user nodes wish to have access to the latest possible version of information, and have a preference for reliable information, i.e., information that originated at the reliable source. Let $S_i(t)$ indicate the reliability status of the information packet present at node $i$ at time $t$, with $S_i(t)=0$ and $S_i(t)=1$ corresponding to reliable and unreliable packet, respectively. At time $t$, if node $i$ sends update to node $j$, node $j$ makes a decision to accept or reject the packet in accordance with the following set of rules:
\begin{itemize}
    \item If $S_i(t)=1$ and $S_j(t)=1$, i.e., both nodes possess unreliable information, then node $j$ chooses the packet with lower version age of information.
    \item If $S_i(t)=0$ and $S_j(t)=0$, i.e., both nodes possess reliable information, then node $j$ again chooses the packet with lower version age of information.
    \item If $S_i(t)=0$ and $S_j(t)=1$, i.e., incoming packet has reliable information but node $j$ has unreliable information, then node $j$ will choose the reliable incoming packet as long as $X_i \leq X_j+1$, in other words, the incoming packet is no more than one version older than the packet already present at node $j$.
    \item If $S_i(t)=1$ and $S_j(t)=0$, i.e., node $j$ already has reliable information packet and incoming packet is unreliable, then node $j$ would continue to keep its reliable packet as long as $X_j \leq X_i+1$, in other words, the reliable packet present at node $j$ is no more than one version older than the incoming unreliable packet.
\end{itemize}

We first study what fraction of users nodes on average in the network have unreliable information when the network exchanges files according to the above protocol, and how it affects the average version age of information at user nodes. In this respect, we model the problem as a stochastic hybrid system. We also study a different setting where user nodes give higher priority to freshness of information and always choose the packet with the lowest version age of information. The reliability status of packets in the latter case would only become relevant when two packets being compared have the same version age but different reliability status, in which case the reliable packet would be chosen. We study the dependency of our results on various network parameters and also compare the results of both the settings. We finally present simulations results to verify our theoretical results and give further insights.

\section{System Model and SHS Characterization}\label{sec:system_mod}

The system model consists of a reliable source ($R$) and an unreliable source ($U$) that transmit updates to a set of $n$ user nodes $\mathcal{N}=\{1,\ldots,n\}$ about a process or event (E) that gets updated according to a Poisson process with rate $\lambda_E$, see Fig.~\ref{fig:system_model}. Both reliable and unreliable sources are assumed to always have access to the latest possible version of information and they send updates to user node $i \in \mathcal{N}$ with rate $\frac{\lambda_R}{n}$ and $\frac{\lambda_U}{n}$, respectively. Further for every $i,j \in \mathcal{N}$, node $i$ sends updates to node $j$ according to a Poisson process with rate $\lambda_{ij}=\frac{\lambda}{n-1}$ as a part of the underlying gossip network. Let $F(t)$ denote the fraction of user nodes that have unreliable information packet at time $t$, then 
\begin{align}\label{eqn:F(t)_def}
    F(t)= \frac{S_1(t)+ S_2(t)+ \ldots + S_n(t)}{n}
\end{align}
since $S_i(t)=1$ indicates that node $i$ has unreliable information. We are interested in characterizing the long-term expectations $F= \lim_{t \to \infty} \mathbb{E}[F(t)]$ and $x_i= \lim_{t \to \infty} \mathbb{E}[X_i(t)]$, $i \in \mathcal{N}$, when nodes accept packets according to the protocol stated in Section~\ref{sec:intro}. Note that the reliability status and version age processes at all user nodes are statistically identical, and consequently $x_1=\ldots=x_n$ and $F=s_1$, where $s_i= \lim_{t \to \infty} \mathbb{E}[S_i(t)]$. Next we use SHS modelling \cite{hespanhashs} to obtain a set of linear equations to derive $s_1$ and $x_1$.

In this regard, we select the continuous state for our SHS model as $(\pmb{S}(t),\pmb{X}(t))\in \mathbb{R}^{2n}$, where $\pmb{S}(t)=[S_1(t),\ldots,S_n(t)]$ and $\pmb{X}(t)=[X_1(t),\ldots,X_n(t)]$ denote the instantaneous reliability status and instantaneous version age, respectively, at the $n$ user nodes at time $t$. We say that transition $(i,j)$ takes place when node $i$ sends an update packet to node $j$, with $(E,E)$ representing event update, and $(U,i)$ and $(R,i)$ representing updates to user node $i$ from unreliable and reliable source, respectively. Since $S_i(t)$ and $X_i(t)$ do not change between transitions, the SHS model operates in a single discrete state, where the continuous state obeys the differential equation $(\pmb{\dot S}(t),\pmb{\dot X}(t))=\pmb{0}_{2n}$.  The set of transitions is 
\begin{align}\label{eqn:L-set_of_transitions}
    \mathcal{L}= \{&(E,E)\} \cup \{(U,i):i \in \mathcal{N}\} \cup \{(R,i):i \in \mathcal{N}\}\nonumber \\
    &\cup \{(i,j):i,j \in \mathcal{N}\}
\end{align}
such that the transition $(i,j)$ resets the state $(\pmb{S},\pmb{X})$  at time $t$ to $\phi_{i,j}(\pmb{S},\pmb{X},t)\in \mathbb{R}^{2n}$ post transition. The rates $\lambda_{ij}$ for each transition $(i,j)$ are given as
\begin{align} \label{eqn:rates_lambda}
\lambda_{ij} = \begin{cases} 
\frac{\lambda_U}{n}, & i=U,j\in \mathcal{N}\\
\frac{\lambda_R}{n}, & i=R,j\in \mathcal{N}\\
\frac{\lambda}{n-1}, & i,j\in \mathcal{N}\\
\lambda_E, & i=E,j=E
\end{cases}
\end{align}

Next we define some variables that will come in handy later. Consider a set $A$ of nodes. Some nodes in this set might have reliable information and others might have unreliable information. Let $R(A) \subseteq A$ denote the subset of nodes that posses reliable information and let $U(A) \subseteq A$ denote the subset of nodes that possess unreliable information. For a continuous state $(\pmb{S},\pmb{X})$ and set of nodes $A$, we define version age of set $A$, $X_A$ as follows:
\begin{itemize}
    \item If $A=\varnothing$, then $X_A=\infty$.
    \item  If $A=R(A)$ or $A=U(A)$, then $X_{\arg \min_{j\in A}X_j}$.
    \item If $X_{\arg \min_{j\in R(A)}X_j} \leq X_{\arg \min_{j\in U(A)}X_j} + 1$, then $X_A= X_{\arg \min_{j\in R(A)}X_j}$.
    \item If $X_{\arg \min_{j\in U(A)}X_j} \leq X_{\arg \min_{j\in R(A)}X_j} -2 $, then $X_A= X_{\arg \min_{j\in U(A)}X_j}$.
\end{itemize}
 Next, we define reliability status of set $A$, $S_A$ as follows:
\begin{itemize}
    \item If $X_{R(A)} \leq X_{U(A)} + 1$, then $S_A= S_{\arg \min_{j\in R(A)}X_j}$.
    \item If $X_{U(A)} \leq X_{R(A)} -2 $, then $S_A= S_{\arg \min_{j\in U(A)}X_j}$.
\end{itemize}
That is, as long as the latest reliable packet is no more than one version older than latest unreliable packet in the set of nodes, the node with the latest reliable packet determines $X_A$ and $S_A$. With this definition, based on transition $(i,j)$ at time $t$, the reset map to $\phi_{i,j}(\pmb{S},\pmb{X},t)=[S_1',\ldots,S_n',X_1',\ldots,X_n']\in \mathbb{R}^{2n}$ can be described as
\begin{align}\label{eqn:continuous_state_Sl}
S_{\ell}' = \begin{cases} 
S_{\{U,\ell\}}, & i=U,j\in \mathcal{N},\ell=j\\
0, & i=R,j\in \mathcal{N},\ell=j\\
S_{\{i,\ell\}}, & i,j\in \mathcal{N},\ell=j\\
S_{\ell}, & \text{otherwise}
\end{cases}
\end{align}
and
\begin{align} \label{eqn:continuous_state_Xl}
X_{\ell}' = \begin{cases} 
X_{\ell}+1, & i=E,j=E,\ell=j\\
\mathbbm{1}_{\{ X_{R(\{\ell\})}=1 \} }, & i=U,j\in \mathcal{N},\ell=j\\
0, & i=R,j\in \mathcal{N},\ell=j\\
X_{\{i,\ell\}}, & i,j\in \mathcal{N}_R,\ell=j\\
X_{\ell}, & \text{otherwise}
\end{cases}
\end{align}
where $\mathbbm{1}_{\{.\}}$ represents the indicator function. Next, we will pick a series of test functions $\psi:\mathbb{R}^{2n}\times [0,\infty) \to \mathbb{R}$ that are time-invariant, i.e., their partial derivative with respect to $t$ is $\frac{\partial\psi(\pmb{X},\pmb{U},t)}{\partial t}=0$, such that their long-term expected value $\mathbb{E}[\psi]=\lim_{t \to \infty} \mathbb{E}[\psi(\pmb{S}(t),\pmb{X}(t),t)]$ will be useful for analysis later. Since the test function only depends on the continuous state values $(\pmb{S},\pmb{X})$ and is time-invariant, for simplicity, we will drop the third input $t$ and write $\psi(\pmb{S},\pmb{X},t)$ as $\psi(\pmb{S},\pmb{X})$, which is assumed to satisfy $\dot \psi(\pmb{S}(t),\pmb{X}(t))=0$. 
Defining $\mathbb{E}[\psi(\phi_{i,j})]=\lim_{t \to \infty} \mathbb{E}[\psi(\phi_{i,j}(\pmb{S}(t),\pmb{X}(t),t))]$, \cite[Thm.~1]{hespanhashs} yields 
\begin{align} \label{eqn:hespanha_eqn}
    0=\sum_{(i,j)\in \mathcal{L}}(\mathbb{E}[\psi(\phi_{i,j})]- \mathbb{E}[\psi] )\lambda_{ij}
\end{align}
which is similar to derivations in \cite{Yates21gossip_traditional,Yates21gossip,kaswan22timestomping}, where the left side is set to zero due to $\frac{d\mathbb{E}[\psi(\pmb{S}(t),\pmb{X}(t),t)] }{dt} =0$ at large $t$ as the expectation stabilizes. We will be using this equation repeatedly by defining a series of time-invariant test functions appropriate for our analysis. For more details, the reader is encouraged to look at references \cite{hespanhashs} and \cite{Yates21gossip_traditional}.

\section{Reliability and Version Age Analysis} \label{sec:analysis}

Since the version age and reliability status evolution processes are identical for all user nodes, we use $A_k$ to denote the typical set of $k$ user nodes. Our first test function is $\psi(\pmb{S},\pmb{X})= S_{A_k}$, which is modified upon transition $(i,j)$ to  $\psi(\phi_{i,j}(\pmb{S},\pmb{X},t))=S_{A_k}'$ and can be characterized using (\ref{eqn:continuous_state_Sl}), (\ref{eqn:continuous_state_Xl}) as follows,
\begin{align} \label{eqn:testfunc_resetmap_SAk}
S_{A_k}' = \begin{cases} 
S_{A_k\cup\{U\}}, & i=U,j\in A_k\\
0, & i=R,j\in A_k\\
S_{A_{k+1}}, & i=\mathcal{N} \backslash A_k,j\in A_k\\
S_{A_k}, & \text{otherwise}
\end{cases}
\end{align}
Defining $a_k= \lim_{t \to \infty} \mathbb{E}[S_{A_k}(t)]$ and $b_k= \lim_{t \to \infty} \mathbb{E}[S_{A_k\cup\{U\}}(t)]$, and using (\ref{eqn:hespanha_eqn}) gives,
\begin{align} \label{eqn:hesp_eqn_a_k}
    0=&(b_k- a_k)\frac{k\lambda_U}{n} +(0-a_k) \frac{k\lambda_R}{n}      \nonumber\\ 
    &+ (a_{k+1} - a_k)\frac{k(n-k)\lambda}{n-1}
\end{align}
Next, we study the test function $\psi(\pmb{S},\pmb{X})= S_{A_k\cup\{U\}}$, which has $(i,j)$ transition map as follows,
\begin{align} \label{eqn:testfunc_resetmap_SAkU}
S_{A_k\cup\{U\}}' = \begin{cases} 
1-\mathbbm{1}_{\{ X_{R(A_k)}=0 \} } , & i=E,j=E\\
0, & i=R,j\in A_k\\
S_{A_{k+1}\cup\{U\}}, & i=\mathcal{N} \backslash A_k,j\in A_k\\
S_{A_k\cup\{U\}}, & \text{otherwise}
\end{cases}
\end{align}
Defining $c_k=\lim_{t \to \infty} \mathbb{E}[\mathbbm{1}_{\{ X_{R(A_k)}(t)=0 \} }]$, (\ref{eqn:hespanha_eqn}) gives,
\begin{align}\label{eqn:hesp_eqn_b_k}
    0=&(1-c_k - b_k)\lambda_E + (0-b_k)\frac{k\lambda_R}{n}  \nonumber\\ 
    &+(b_{k+1}-b_k)\frac{k(n-k)\lambda}{n-1} 
\end{align}
Finally, we study the test function $\psi(\pmb{S},\pmb{X})= \mathbbm{1}_{\{ X_{R(A_k)}=0 \} }$, which has the $(i,j)$ transition map as follows,
\begin{align} \label{eqn:testfunc_resetmap_1(XRA=0)}
\mathbbm{1}_{\{ X_{R(A_k)}=0 \} }' = \begin{cases} 
0, & i=E,j=E\\
1, & i=R,j\in A_k\\
\mathbbm{1}_{\{ X_{R(A_{k+1})}=0 \} }, & i=\mathcal{N} \backslash A_k,j\in A_k\\
\mathbbm{1}_{\{ X_{R(A_k)}=0 \} }, & \text{otherwise}
\end{cases}
\end{align}
that, upon employing (\ref{eqn:hespanha_eqn}), gives,
\begin{align}\label{eqn:hesp_eqn_c_k}
    0=&(1-c_k)\lambda_E + (1-c_k)\frac{k\lambda_R}{n} \nonumber\\
    &+ (c_{k+1}-c_k)\frac{k(n-k)\lambda}{n-1}
\end{align}
Equations (\ref{eqn:hesp_eqn_a_k}), (\ref{eqn:hesp_eqn_b_k}), (\ref{eqn:hesp_eqn_c_k}) can be rewritten as follows,
\begin{align}
    a_k&= \frac{b_k\frac{k\lambda_U}{n} 
    + a_{k+1}\frac{k(n-k)\lambda}{n-1}}{\frac{k\lambda_U}{n} + \frac{k\lambda_R}{n} +  \frac{k(n-k)\lambda}{n-1}} \label{eqn:formula_a_k}\\
    b_k&= \frac{(1-c_k )\lambda_E 
    +b_{k+1}\frac{k(n-k)\lambda}{n-1} }{\lambda_E + \frac{k\lambda_R}{n} + \frac{k(n-k)\lambda}{n-1}  } \label{eqn:formula_b_k}\\
    c_k&=\frac{  \frac{k\lambda_R}{n} + c_{k+1}\frac{k(n-k)\lambda}{n-1} }{ \lambda_E +   \frac{k\lambda_R}{n} +  \frac{k(n-k)\lambda}{n-1} } \label{eqn:formula_c_k}
\end{align}

Note that $a_1=s_1=F$, and therefore, for computation of $F$, we first solve for $c_k$ by (\ref{eqn:formula_c_k}), starting from $k=n$ and successively substituting for $k=n-1,\ldots,1$ in an iterative fashion. Once we have computed all $c_k$, we can similarly backward iterate on (\ref{eqn:formula_b_k}) to compute all the $b_k$. Finally, we can use $b_k$ and the inductive equation (\ref{eqn:formula_a_k}) to compute all the $a_k$, starting from $a_n$ to reach $a_1=F$ at the end. 

Couple of observations can be made from equations (\ref{eqn:formula_a_k}), (\ref{eqn:formula_b_k}), (\ref{eqn:formula_c_k}). First if $\lambda_R$ is large, then substituting $\lambda_R = \infty $ in (\ref{eqn:formula_a_k}) gives $a_1=0$, i.e., all nodes have reliable information at all times. On the other hand, if $\lambda_U \to \infty$, then $a_1 \approx b_1$, where $b_1$ is completely independent of $\lambda_U$ and is a positive quantity depending on other transition rates and network size $n$. Hence when $\lambda_U$ is large, a non-zero fraction of nodes are still expected to have reliable information, this can be attributed to the fact that nodes have a preference for reliable information even if it amounts to reverting to an older version of information, and as long as $\lambda_R$ is non-zero, some nodes are successfully able to discard their unreliable packets in favor of reliable packets. Further, if the gossiping rate $\lambda$ is high, then $a_1 \approx \ldots \approx a_n$, which results in
\begin{align} \label{eqn:analysis_large_lambda_gossiping}
    a_1 &\approx a_n \approx  b_n \frac{\lambda_U}{\lambda_U+ \lambda_R} \approx \frac{(1-c_n)\lambda_E}{\lambda_E+\lambda_R} \frac{\lambda_U}{\lambda_U+ \lambda_R}\nonumber\\
        &\approx \left( \frac{\lambda_E}{\lambda_E+\lambda_R} \right)^2\frac{\lambda_U}{\lambda_U+ \lambda_R}
\end{align}
One might expect that a high gossiping rate would result in fast dissemination of fresh reliable information in the network and push $F$ to $0$, however, as seen in (\ref{eqn:analysis_large_lambda_gossiping}), $F$ in this case would still depend on other things such how fast the reliable source can start sending packets soon after the event gets updated and if unreliable source is quicker in this regard.

Next, we characterize the version age at the user nodes. We define $e_k=\lim_{t \to \infty} \mathbb{E}[X_{A_k}(t)]$ and $d_k=\lim_{t \to \infty} \mathbb{E}[\mathbbm{1}_{\{ X_{R(A_k)}(t)=1 \} }]$. We require two new test functions $X_{A_k}$ and $\mathbbm{1}_{\{ X_{R(A_k)}}$, that are modified by transition $(i,j)$ as follows,

\begin{align} \label{eqn:testfunc_resetmap_XAk}
X_{A_k}' = \begin{cases} 
X_{A_k}+1, & i=E,j=E\\
\mathbbm{1}_{\{ X_{R(A_k)}=1 \} }, & i=U,j\in A_k\\
0, & i=R,j\in A_k\\
X_{A_{k+1}}, & i=\mathcal{N} \backslash A_k,j\in A_k\\
X_{A_k}, & \text{otherwise}
\end{cases}
\end{align}

\begin{align} \label{eqn:testfunc_resetmap_1XRAk=1}
\mathbbm{1}_{\{ X_{R(A_k)}=1 \} }' = \begin{cases} 
\mathbbm{1}_{\{ X_{R(A_k)}=0 \} } , & i=E,j=E\\
0, & i=R,j\in A_k\\
\mathbbm{1}_{\{ X_{R(A_{k+1})}=1 \} }, & i=\mathcal{N} \backslash A_k,j\in A_k\\
S_{A_k\cup\{U\}}, & \text{otherwise}
\end{cases}
\end{align}

Using (\ref{eqn:hespanha_eqn}), they give the linear equations, 
\begin{align}\label{eqn:hesp_eqn_e_k}
    0=&(e_k+1-e_k)\lambda_E+ (d_k- e_k)\frac{k\lambda_U}{n} + (0-e_k) \frac{k\lambda_R}{n}  \nonumber\\ 
    &+ (q_{k+1} - e_k)\frac{k(n-k)\lambda}{n-1} 
\end{align}
\begin{align}\label{eqn:hesp_eqn_d_k}
    0=&(c_k - d_k)\lambda_E + (0-d_k)\frac{k\lambda_R}{n}    \nonumber\\ 
    &+(d_{k+1}-d_k)\frac{k(n-k)\lambda}{n-1}   =0
\end{align}
which upon rearrangement, along with (\ref{eqn:formula_c_k}), give us the following set of equations,
\begin{align}
    e_k &= \frac{\lambda_E+ d_k\frac{k\lambda_U}{n} 
    + e_{k+1} \frac{k(n-k)\lambda}{n-1}}{\frac{k\lambda_U}{n} + \frac{k\lambda_R}{n} +  \frac{k(n-k)\lambda}{n-1}} \label{eqn:formula_e_k}\\
    d_k &= \frac{c_k\lambda_E 
    +d_{k+1}\frac{k(n-k)\lambda}{n-1} }{\lambda_E + \frac{k\lambda_R}{n} + \frac{k(n-k)\lambda}{n-1}  } \label{eqn:formula_d_k}\\
    c_k &=\frac{  \frac{k\lambda_R}{n} + c_{k+1}\frac{k(n-k)\lambda}{n-1} }{ \lambda_E +   \frac{k\lambda_R}{n} +  \frac{k(n-k)\lambda}{n-1} } \label{eqn:formula_c_k2}
\end{align}
As before, we first compute all $c_k$ starting from $k=n$, which next allows us to compute all $d_k$ iteratively, which in turn allows us to compute all $e_k$, to finally give us version age $x_1=e_1$. Again if $\lambda_R$ is large, it leads to $e_1 \approx 0$, as the network is flooded with fresh reliable packets. On the other hand if $\lambda_U \to \infty$, then $e_1 \approx d_1$, where $d_k$ is a positive quantity independent of $\lambda_U$. That is, version age of information of user nodes is not guaranteed to drop to zero when unreliable source transmits updates at high rate, since a reliable packet of age one is more preferred than a zero age unreliable packet. 

\section{Higher Preference for Fresh Information}\label{sec:freshness}
For further perspective, we now study the case where the user nodes gives highest priority to fresh information. This means when a user node receives a packet, its primary goal is to keep the packet which has the lowest version age of information and only when the two packets have the same version age but different reliability status will the node prefer the reliable packet over the unreliable packet. Considering the network topology and transition rates to be the same as in Section~\ref{sec:system_mod}, we modify the definition of $X_A$ as follows:
\begin{itemize}
    \item If $A=\varnothing$, then $X_A=\infty$, else $X_A= X_{\arg \min_{j\in A}X_j}$.
\end{itemize}
We modify the definition $S_A$ as follows:
\begin{itemize}
    \item If $X_{R(A)} \leq X_{U(A)}$, then $S_A= S_{\arg \min_{j\in R(A)}X_j}$, else $S_A= S_{\arg \min_{j\in U(A)}X_j}$.
\end{itemize}
The test functions of interest are $S_{A_k}$ and $S_{A_k\cup\{U\}}$ with $(i,j)$ transition maps,
\begin{align} \label{eqn:testfunc_resetmap_SAk_freshpriority}
S_{A_k}' = \begin{cases} 
S_{A_k\cup\{U\}}, & i=U,j\in A_k\\
0, & i=R,j\in A_k\\
S_{A_{k+1}}, & i=\mathcal{N} \backslash A_k,j\in A_k\\
S_{A_k}, & \text{otherwise}
\end{cases}
\end{align}
and
\begin{align} \label{eqn:testfunc_resetmap_SAkU_freshpriority}
S_{A_k\cup\{U\}}' = \begin{cases} 
1 , & i=E,j=E\\
0, & i=R,j\in A_k\\
S_{A_{k+1}\cup\{U\}}, & i=\mathcal{N} \backslash A_k,j\in A_k\\
S_{A_k\cup\{U\}}, & \text{otherwise}
\end{cases}
\end{align}
Defining $\bar{a}_k= \lim_{t \to \infty} \mathbb{E}[S_{A_k}(t)]$ and $\bar{b}_k= \lim_{t \to \infty} \mathbb{E}[S_{A_k\cup\{U\}}(t)]$, and employing (\ref{eqn:hespanha_eqn}) for the above test functions gives
\begin{align}
    \bar{a}_k  &= \frac{\bar{b}_k\frac{k\lambda_U}{n} 
    + \bar{a}_{k+1}\frac{k(n-k)\lambda}{n-1}}{\frac{k\lambda_U}{n} + \frac{k\lambda_R}{n} +  \frac{k(n-k)\lambda}{n-1}} \label{eqn:formula_bar_a_k} \\
    \bar{b}_k  &= \frac{\lambda_E 
    +(\bar{b}_{k+1})\frac{k(n-k)\lambda}{n-1} }{\lambda_E + \frac{k\lambda_R}{n} + \frac{k(n-k)\lambda}{n-1}  } \label{eqn:formula_bar_b_k}
\end{align}
 Computing version age at user nodes in this case is simpler with only one test function required $X_{A_k}$ with $\bar{e}_k=\lim_{t \to \infty} \mathbb{E}[X_{A_k}(t)]$. The corresponding transition map and linear equation is
 \begin{align} \label{eqn:testfunc_resetmap_XAk_freshpriority}
X_{A_k}' = \begin{cases} 
\bar{X}_{A_k}+1, & i=E,j=E\\
0, & i=U,j\in A_k\\
0, & i=R,j\in A_k\\
X_{A_{k+1}}, & i=\mathcal{N} \backslash A_k,j\in A_k\\
X_{A_k}, & \text{otherwise}
\end{cases}
\end{align}
and
\begin{align}\label{eqn:formula_bar_e_k}
    \bar{e}_k= \frac{\lambda_E+ \bar{e}_{k+1} \frac{k(n-k)\lambda}{n-1}}{\frac{k\lambda_U}{n} + \frac{k\lambda_R}{n} +  \frac{k(n-k)\lambda}{n-1}}
\end{align}

Comparing $\bar{e}_k$ and $\bar{a}_k$ with $e_k$ and $a_k$ of the previous setting, we can see iteratively starting from $k=n$, that $\bar{e}_k \geq e_k$ and $\bar{b}_k \leq b_k$ leading to $\bar{a}_k \leq a_k$. Thus there is a tradeoff between information reliability and version age. However, when $\lambda_E$ is large, then $d_k \to 0$ and $c_k \to 0$ from (\ref{eqn:formula_d_k}) and (\ref{eqn:formula_c_k2}), and hence $\bar{e}_k \approx e_k$ for large $\lambda_E$. This is because the event is getting updated so fast that all packets in the network are quickly getting outdated, leaving no reliable packets with a small version age of one in the network, making $d_k \approx 0$.

\section{Numerical Results}\label{sec:num_results}

\begin{figure}[t]
 	\begin{center}
 	\subfigure[]{\includegraphics[width=0.49\linewidth]{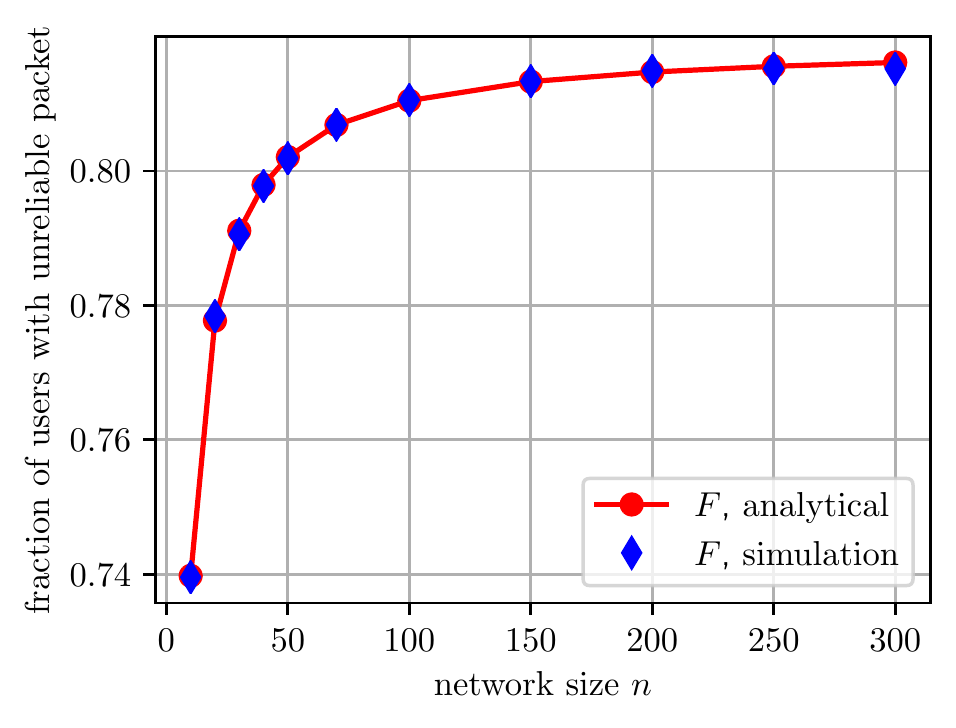}}
 	\subfigure[]{\includegraphics[width=0.477\linewidth]{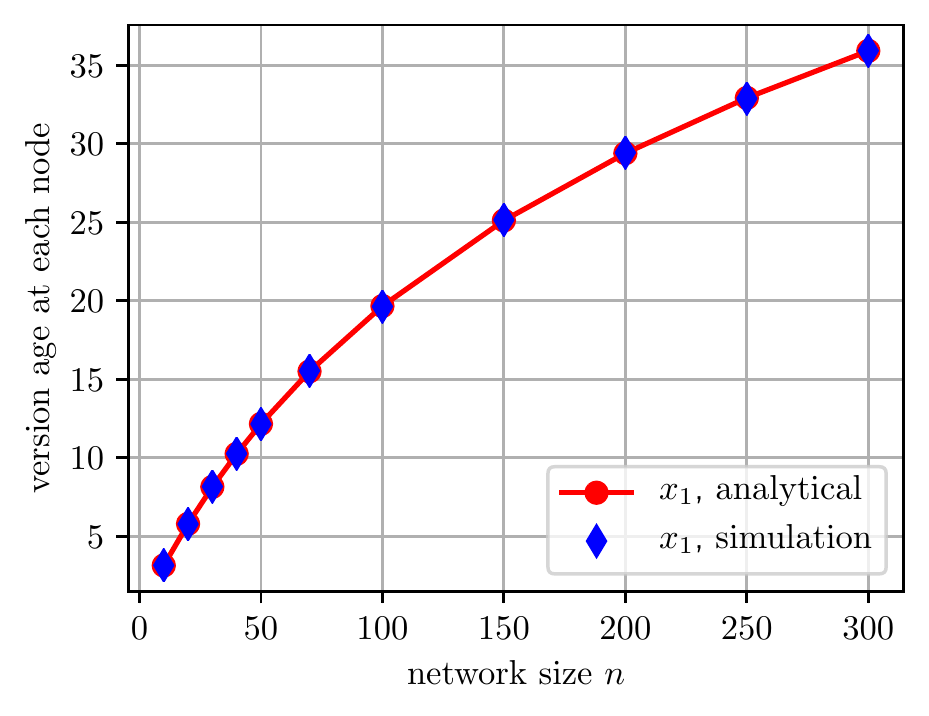}}
 	\end{center}
 	\vspace{-0.4cm}
 	\caption{Analytical and simulation results compared for (a) expected fraction of users with unreliable information $F$. (b) expected version age $x_1$.}
 	\label{fig:simulation_vs_formula}
 	\vspace{-0.5cm}
 \end{figure}

We simulate a fully connected network for various network sizes $n$ with parameters $\lambda_E=2$, $\lambda_U=5$, $\lambda_R=1$ and $\lambda=0.1$ for upto a total time of $10^6$ which we use as proxy for $t \to \infty$ and plot simulation points (blue dots) of $F$ and $x_1$ on curves (red lines) obtained from equations (\ref{eqn:formula_a_k}), (\ref{eqn:formula_b_k}), (\ref{eqn:formula_c_k}), (\ref{eqn:formula_e_k}), (\ref{eqn:formula_d_k}) in Fig.~\ref{fig:simulation_vs_formula}. The real-time simulation points coincide with the iterative calculation of the derived equation curves, lending support to the theoretical results and show that both $F$ and $x_1$ increase with network size $n$.

\begin{figure}[t]
 	\begin{center}
 	\subfigure[]{\includegraphics[width=0.48\linewidth]{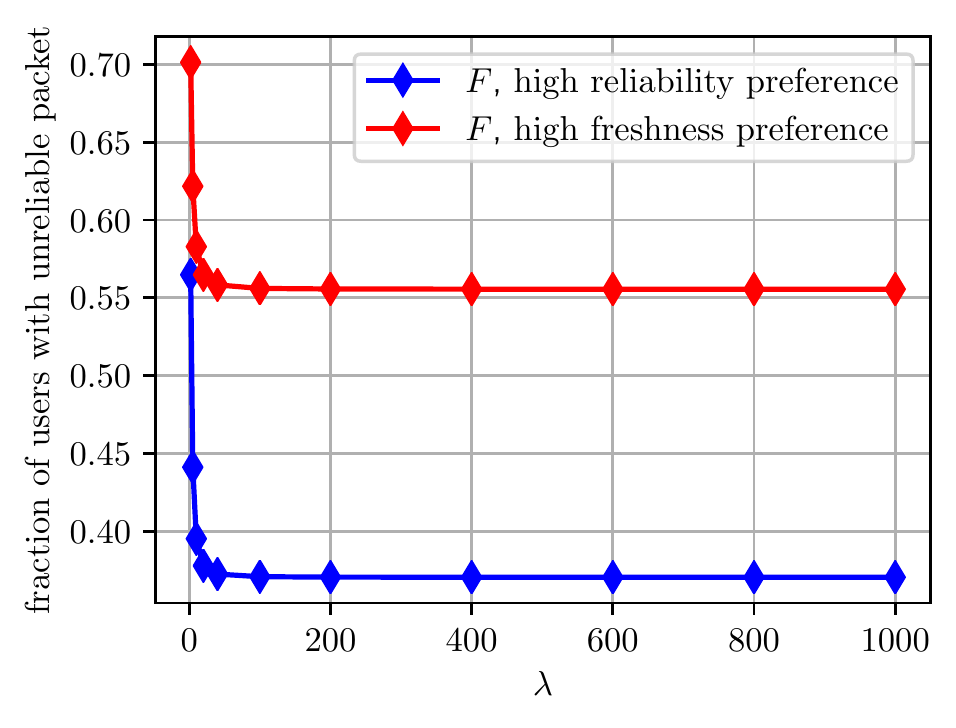}}
  \subfigure[]{\includegraphics[width=0.49\linewidth]{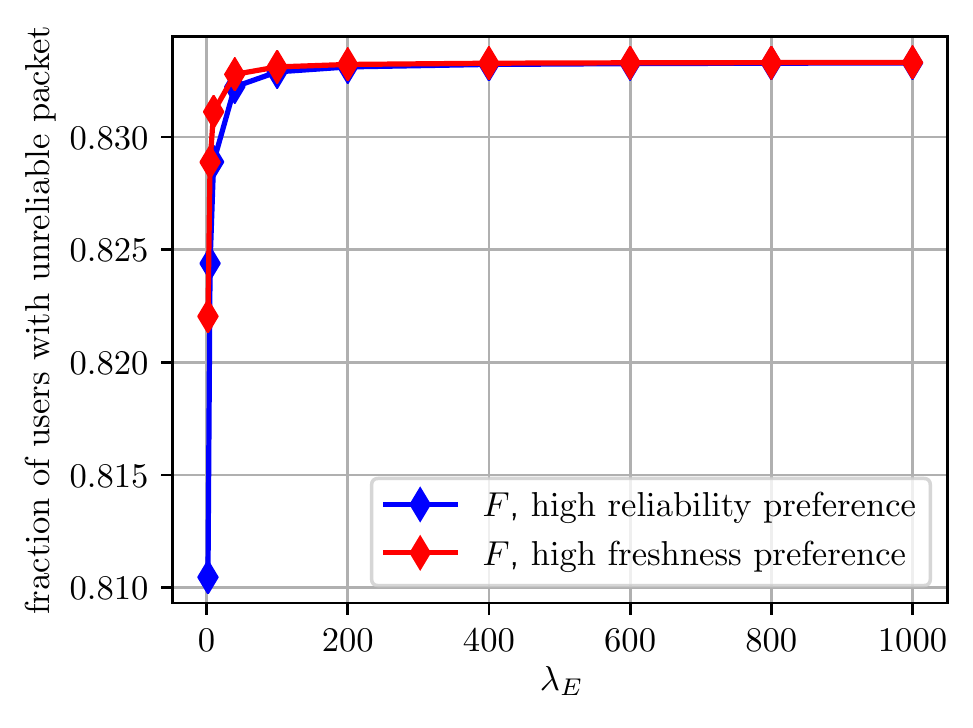}}
 	\subfigure[]{\includegraphics[width=0.47\linewidth]{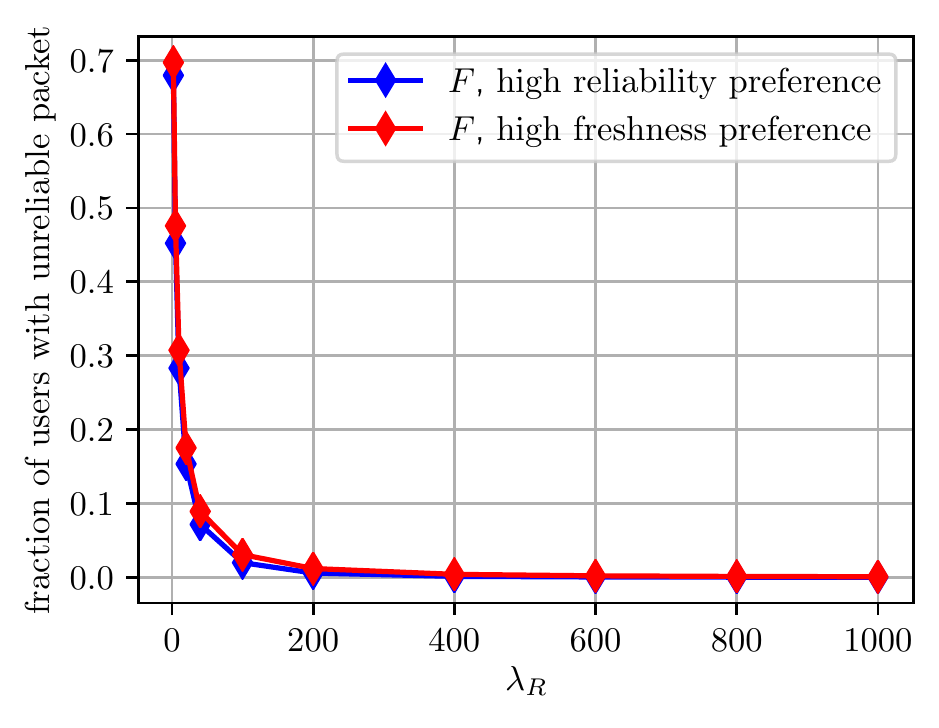}}
  \subfigure[]{\includegraphics[width=0.49\linewidth]{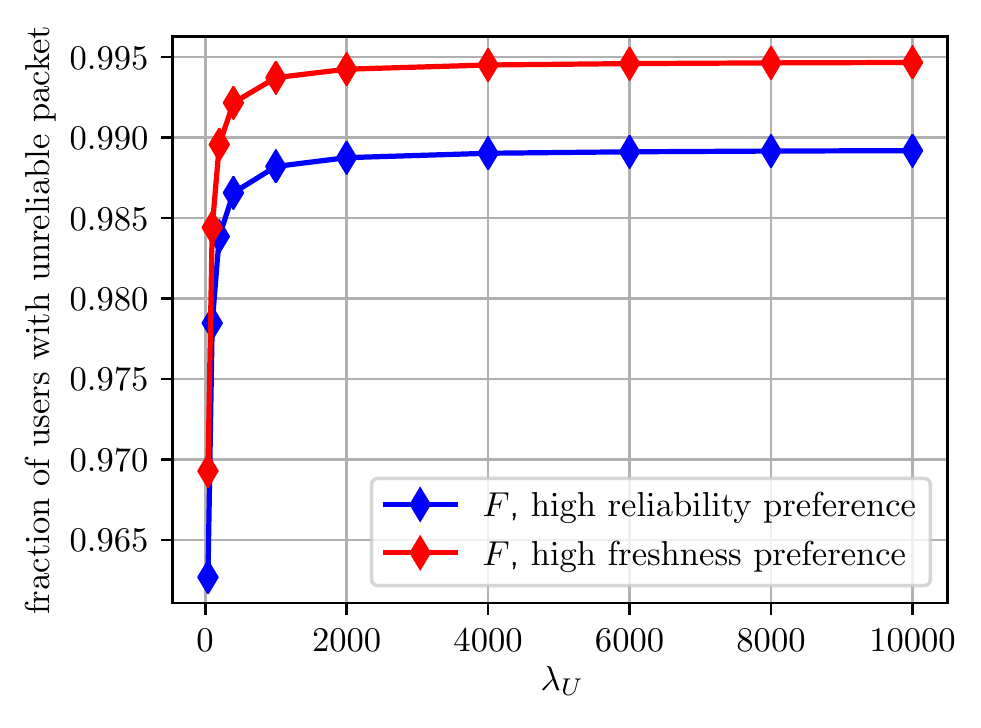}}
 	\end{center}
 	\vspace{-0.4cm}
 	\caption{ $F$ as a function of network parameters $\lambda$, $\lambda_E$, $\lambda_R$ and $\lambda_U$.}	
  \label{fig:status_vs_rates}
 	\vspace{-0.7cm}
 \end{figure}

\begin{figure}[t]
 	\begin{center}
 	\subfigure[]{\includegraphics[width=0.47\linewidth]{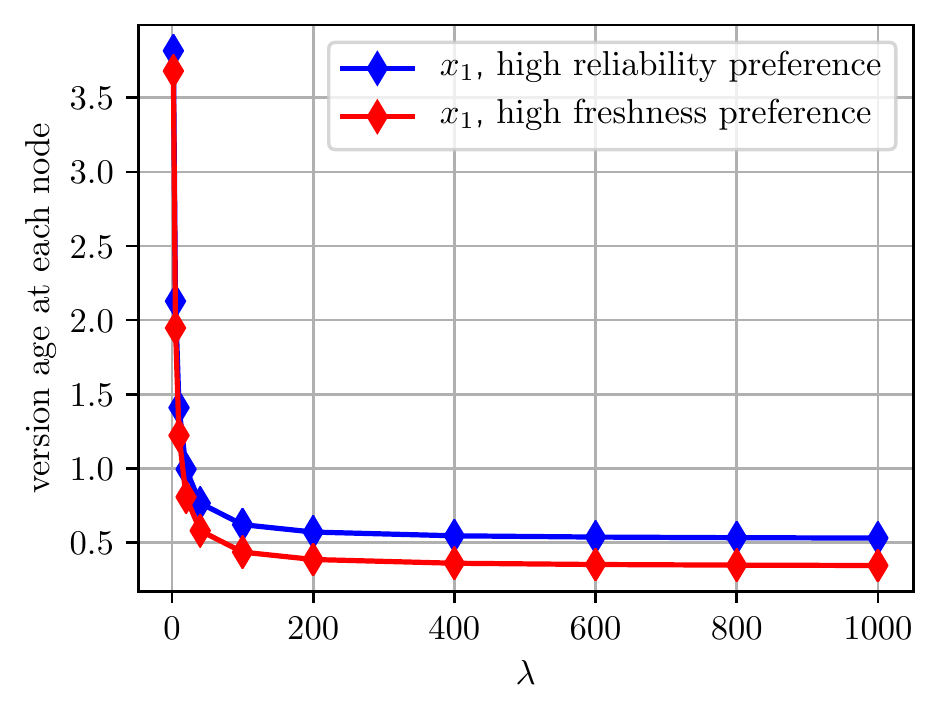}}
  \subfigure[]{\includegraphics[width=0.49\linewidth]{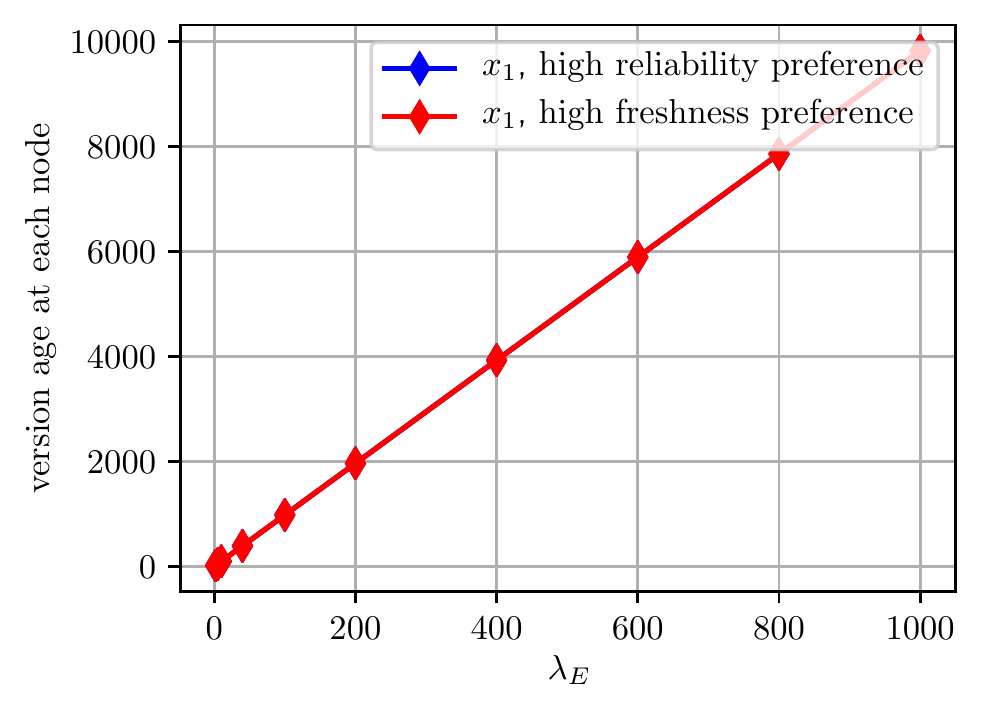}}
 	\subfigure[]{\includegraphics[width=0.48\linewidth]{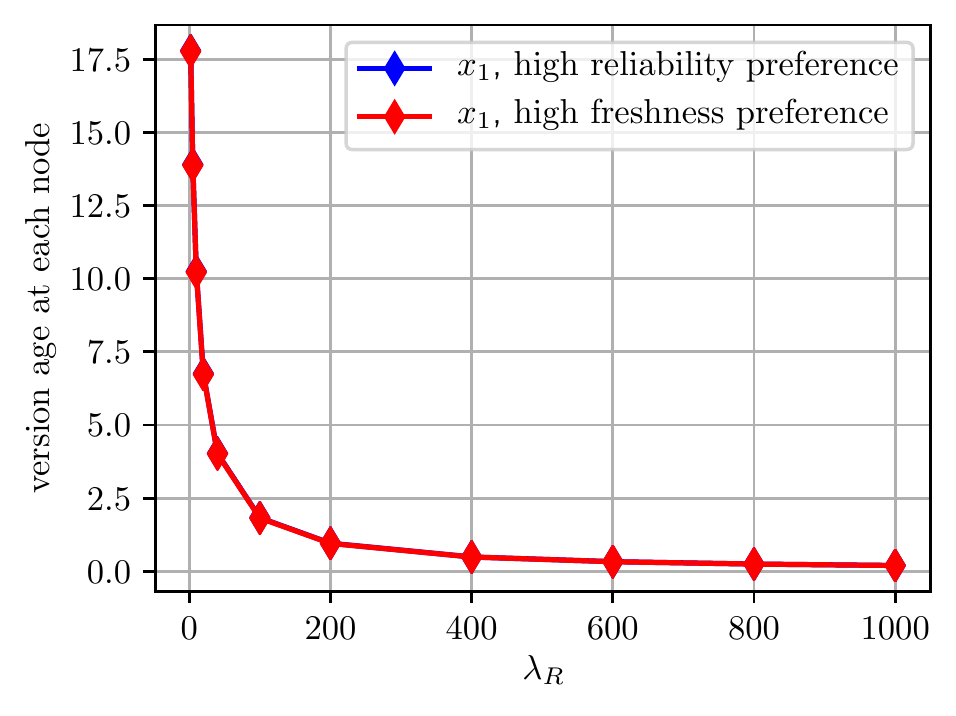}}
  \subfigure[]{\includegraphics[width=0.49\linewidth]{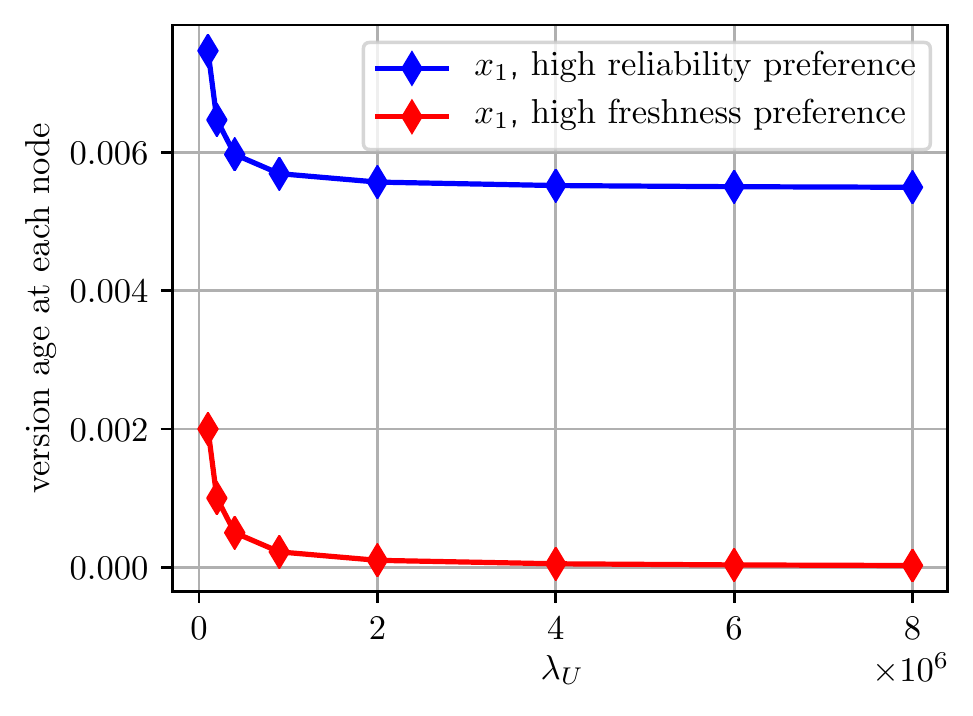}}
 	\end{center}
 	\vspace{-0.4cm}
 	\caption{$x_1$ as a function of network parameters $\lambda$, $\lambda_E$, $\lambda_R$ and $\lambda_U$. }
 	\label{fig:age_vs_rates}
 	\vspace{-0.8cm}
 \end{figure}

We next plot $F$ and $x_1$ for $n=100$ nodes in Fig.~\ref{fig:status_vs_rates} and Fig.~\ref{fig:age_vs_rates}, respectively, as function of the network parameters $\lambda_E$, $\lambda_U$, $\lambda_R$ and $\lambda$. Fig.~\ref{fig:status_vs_rates}(a) shows that $F$ converges to a non-zero value $ \left( \frac{\lambda_E}{\lambda_E+\lambda_R} \right)^2\frac{\lambda_U}{\lambda_U+ \lambda_R}=0.37$ as predicted in (\ref{eqn:analysis_large_lambda_gossiping}). At large $\lambda_R$, $F \to 0$ and $x_1 \to 0$ in  Fig.~\ref{fig:status_vs_rates}(c) and Fig.~\ref{fig:age_vs_rates}(c), respectively, as discussed in Section~\ref{sec:analysis}, whereas a large $\lambda_U$ does not have the same effect where $F$ converges to a value strictly less than one in Fig.~\ref{fig:status_vs_rates}(d). Further from Fig.~\ref{fig:age_vs_rates}(d), we see that the version age does not converge to zero in high reliability preference case as discussed in Section~\ref{sec:analysis}, but it goes to zero in the high freshness case which prioritizes fetching any packet with a potential to decrease its age. Lastly, as $\lambda_E$ becomes large, we see that $x_1$ coincides for both high reliability and high freshness preference cases, as predicted in Section~\ref{sec:freshness}, and scales linearly with $\lambda_E$, since all terms in the iteration (\ref{eqn:formula_bar_e_k}) will have a common factor $\lambda_E$.

\bibliographystyle{unsrt}
\bibliography{ref_priyanka}

\end{document}